# Terahertz-driven magnetism dynamics in the orthoferrite $DyFeO_3$

A. H. M. Reid,[1, a)] Th. Rasing,[3] R. V. Pisarev,[4] H. A. Dürr,[1] and M. C. Hoffmann[2]

[1)]*Stanford Institute for Materials and Energy Science, SLAC National Accelerator Laboratory, 2575 Sand Hill Road, Menlo Park, 94025 CA, United States*

[2)]*LCLS Laser Department, SLAC National Accelerator Laboratory, 2575 Sand Hill Road, Menlo Park, 94025 CA, United States*

[3)]*Radboud University Nijmegen, Institute for Molecules and Materials, Heijendaalseweg 135, 6525 AJ Nijmegen, The Netherlands*

[4)]*Ioffe Physical-Technical Institute, Russian Academy of Sciences, 194021 St. Petersburg, Russia*

Terahertz driven magnetization dynamics are explored in the orthoferrite $DyFeO_3$. A high-field, single cycle THz pulse is used to excite magnon modes in the crystal together with other resonances. Both quasi-ferromagnetic and quasi-antiferromagnetic magnon modes are excited and appear in time-resolved measurements of the Faraday rotation. Other modes are also observed in the measurements of the time-resolved linear birefringence. Analysis of the excitation process reveals that despite larger than expected electro-optical susceptibility it is mainly the THz magnetic field that couples to the quasi-ferromagnetic and quasi-antiferromagnetic magnon branches.

Magnonic technologies aim to use magnetic waves, magnons, as a means to store, transport and operate on data in analogies of electrical circuits. This offers potential advantages over conventional charge-based electronics such as non-volatility, energy efficiency, a lower sensitivity to electrostatic discharge and access to the intrinsic resonances that magnetism provides, while retaining favorable scaling properties [1,2]. However, these technologies are at the very early stages of development with the focus remaining on concept devices using the prototype materials such as permalloy and yttrium iron garnet [2]. Great potential remains in extending these concepts to more advanced magnetic materials where there is the potential to greatly extend the operating frequency and the efficiency.

In this letter we explore methods to coherently control magnetic excitations in $DyFeO_3$. Coherent excitation of magnetic resonances in $DyFeO_3$ have been demonstrated at near-infrared optical frequencies [3,4]. Here torques are applied to the magnetization via non-linear electric field interactions such as the inverse Faraday and inverse Cotton-Mouton effects.

Several groups have recently demonstrated the resonant excitation of magnons using terahertz pulses in similar systems, namely NiO [5] and YFeO$_3$ [6-8]. It this spectral region it has been assumed that the excitation of magnons occurs via the direct magnetic dipole coupling to the electro-magnetic wave and non-linear interactions with the electric field can be ignored. Here we explore this assumption in detail in a prototype system DyFeO$_3$ by using the magnetic field of strong-field single cycle pulses to excite coherent magnon oscillations.

The orthoferrites, $R$FeO$_3$ (where $R$ is typically a rare earth ion), have significant potential in magnetic technologies. The room temperature (300 K) magnetic order is that of weak ferromagnetism; where the magnetic exchange between the spins of Fe$^{+3}$ ions ($L=0$, $S=5/2$) in DyFeO$_3$ is antiferromagnetic in the *ac*-plane, however, with a small magnetization along the *c*-axis due to a canting of the spins in the antisymmetric exchange field [9]. This net magnetization potentially allows magnetic control akin to ferromagnets, while the underlying antiferromagnet order stiffens the magnon modes due to the super-exchange interaction [10] There are four Fe$^{3+}$ spin sites per unit cell, which order antiferromagnetically to form two sublattices **M$_1$** and **M$_2$**, with vectors directions primarily along the *a*-axis in zero field. The net magnetic moment **M** = **M$_1$**+**M$_2$**, while the antiferromagnetic vector is defined as **L** = **M$_1$**−**M$_2$**. At the center of the Brillouin zone magnon frequencies are approximately in the range 0.1 to 0.6 THz by two orders of magnitude higher than in typical ferromagnetic materials [11-13, 3,4]. The magnetic moments of the Dy$^{3+}$ ions ($L=5$, $S=5/2$) remain in a disordered paramagnetic state above 3 K.

The DyFeO$_3$ sample used in this study was a single-crystal platelet, 38 μm in thickness, cut such that the surface normal is the crystallographic *c*-axis. The magnetic easy axis along the out-of-plane *c*-direction, but an external field $B_{ext} = \mu_0 H_{ext}$ = 0.3 T is applied in-plane to torque magnetization slightly into the plane to an angle of approximately 3.6 degrees [14].

Single cycle THz pulses with pulse energies of 3.5 μJ and frequency spectrum spanning the range from 0.1 to 2.5 THz were generated by optical rectification in LiNbO$_3$ using the tilted pulse front method [15]. The THz beam was focused onto the sample to a spot size of 1 mm, yielding peak electric (**E**) field strengths of 34 MV/m and magnetic flux density of 0.11 T, comparable to the external field. The field was characterized in the time domain using electro-optical sampling in a 150 μm thick (110) cut GaP crystal (see Fig. 1(a)). The Fourier spectrum of the THz pump field (inset) shows slight absorption lines from residual water vapour in the beam path. The oscillations in the time domain corresponding to absorption lines below 0.8 THz are weak compared to the peak THz field and do not adversely affect our measurements. We define the pump-probe overlap $t_0$ as the relative time delay at the maximum electro-optical modulation. For the DyFeO$_3$ measurements, the sample is inserted at exactly the same position (within 50 µm precision) as the electro-optical sampling crystal.

The experimental geometry is shown in Fig. 1(b). The THz pump pulse was incident on the $DyFeO_3$ normal to the sample surface with the 800 nm probe pulse collinear to the propagation of the pump. The probe polarization was oriented to correspond to a minimum in the linear birefringence of the sample and the polarization rotation is measured using a polarization splitting prism and a pair of balanced diode detectors. The degree of rotation is calibrated by a static measurement of diodes signals.

Results from a typical time-resolved measurement are shown in Fig. 1b. At the pump–probe overlap time a strong transient appears in the time-resolved rotation. The size of this change is observed to be correlated with the sample's birefringence; its maximum occurs at the same angle as the sample maximum static linear birefringence. The transient is followed by coherent oscillations with a period of about 2 ps. Extrapolation of the phase suggests that the oscillatory contribution to the signal has a minimum at the peak of the THz $E$-field signal (dashed line).

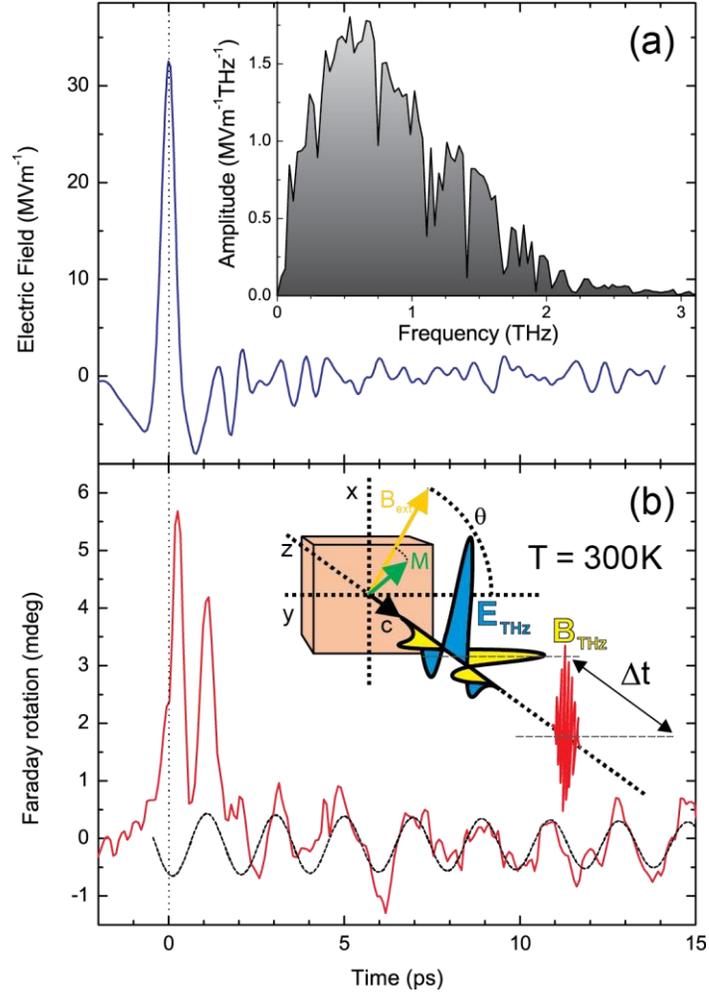

FIG. 1. (Color online) (a) Electro-optical measurement of the THz electric field $E_{THz}$ in a GaP (110) crystal, with the frequency spectrum inset. (b) Time resolved Faraday rotation of an 800 nm probe pulse in DyFeO$_3$ at 300 K following THz pumping, together with the geometry of the experiment. The dashed line is a sine fit to the magnon oscillation at 0.51 THz.

Fig. 2a shows amplitude spectra obtained by Fourier transformation of the time-domain Faraday signals following excitation with the external magnetic field $B_{ext}$ parallel ($\theta=0$) and perpendicular ($\theta=\pi/2$) to the magnetic field component $B_{THz}$ of the THz pulse. Two distinct bands occur. The mode at 0.51 THz corresponds to the quasi-antiferromagnetic (QAFM) resonance, while the band at 0.37 THz is the quasi-ferromagnetic (QFM) mode.

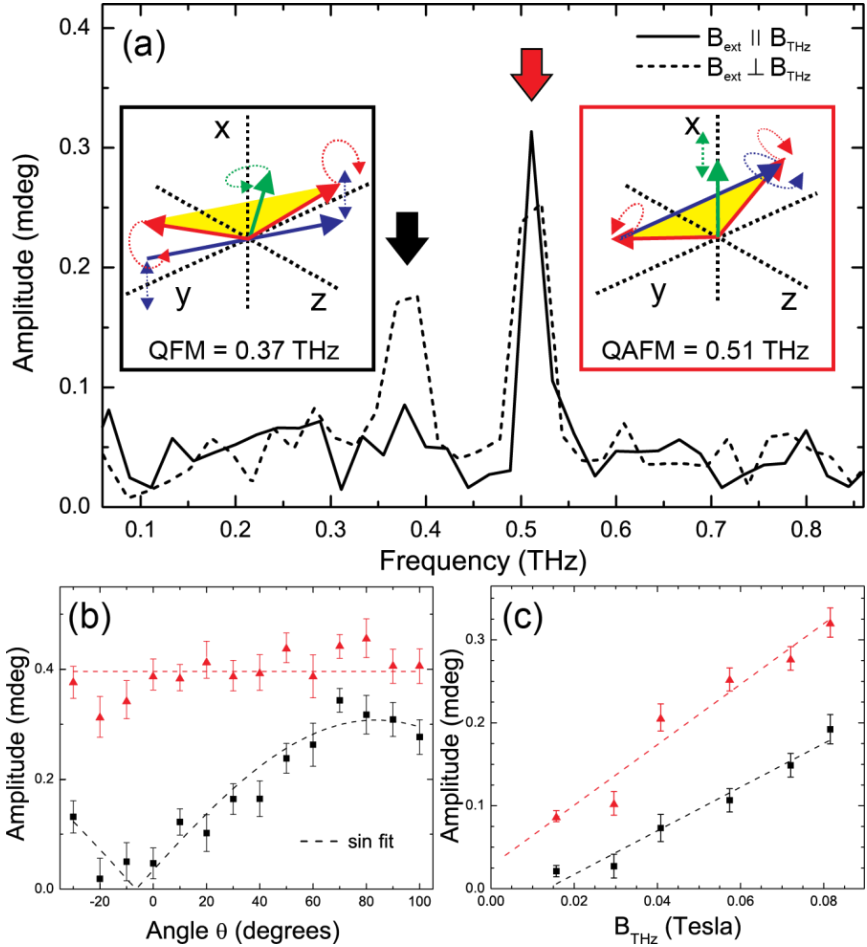

FIG. 2. (Color online) (a) Spectral amplitude of the Faraday with $B_{ext}$ parallel ($\theta=0$), and perpendicular ($\theta=\pi/2$) to $B_{THz}$. Two distinct modes (indicated by arrows) are observed, one at 0.51 THz, corresponding to the quasi-antiferromagnetic resonance, and the other at 0.375 THz, corresponding to quasi-ferromagnetic resonance. (b) The amplitudes of the modes, determined by Fourier transform, as a function of the applied field $B_{ext}$ angle $\theta$. Error bars are determined by spectral amplitude in the 'empty' spectral region from 0.6-0.8 THz. (c) Mode amplitude as a function of THz field strength. Triangles indicate the 0.51 THz mode, squares indicate the 0.375 THz mode.

Measurements were taken of the time-resolved rotation of polarization while the angle of the magnetic field was rotated in the (*ab*) plane of the sample. When the field was applied such that $B_{ext} \parallel B_{THz}$ a single coherent oscillation was observed at 0.51 THz, however, a second oscillation was observed at 0.375 THz as the field was rotated. This mode showed a maximum at an angle of $83^0$ degrees between $B_{ext}$ and $B_{THz}$ (FIG. 2c). The frequencies of the two modes, 0.375 THz and 0.51 THz correspond closely to previous measurements of quasi-ferromagnetic (QFM) and quasi-antiferromagnetic (QAFM) magnon branches at room temperature in $DyFeO_3$ at 0.38 THz and 0.53 THz [11] and 0.38 THz; 0.51 THz [12]. To examine if the oscillations were magnetic in origin time-resolved measurements of the Faraday rotation were compared to measurements of the crystal's linear birefringence (ellipticity) and of the polarization rotation away from a point of minimum birefringence, shown in Fig. 3. While both 0.375 THz and 0.51 THz modes appear in the magneto-optical Faraday signal, neither is present in the measurements of linear birefringence. This supports their identification as the QFM and QAFM magnon branches.

The 0.375 THz QFM mode shows a strong dependence on the relative orientation of the $\mathbf{B_{THz}}$ and external magnetic field $\mathbf{B_{ext}}$, while the 0.51 THz QAFM mode has constant amplitude when the angle θ between $\mathbf{B_{THz}}$ and $\mathbf{B_{ext}}$ is varied. The amplitude dependence of QFM mode shows $\sin^2\theta$ behavior, consistent with the excitation occurring by the magnetic torque, $\boldsymbol{\tau} = \mathbf{M}(\mathbf{B_{ext}}) \times \mathbf{B_{THz}}$. In order to confirm the field dependence of the excitation mechanism, we carried out THz-field dependent measurements. Here the $\mathbf{B_{THz}}$ field strength was continuously attenuated using a pair of wire-grid polarizers, where the first one was rotated while the second one was kept fixed. Linearity of the field attenuation was confirmed using electro-optical sampling at the sample position. Figure 2c shows the amplitudes of the QAFM and QFM resonances as function of $\mathbf{B_{THz}}$-field. The response is linear in $B_{THz}$, demonstrating that the excitation is not driven by an impulsive stimulated Raman scattering, which would scale quadratically with $E_{THz}$ and linear in THz intensity [16].

Several other resonances appear in the optical measurement of the linear birefringence following THz excitation. A total of five distinct modes can be observed in birefringence measurements (FIG. 3), with frequencies of 0.172, 0.205, 0.230, 0.255, and 0.280 THz. These resonances are well below the phonon, magnon and crystal-field frequencies present in $DyFeO_3$ [17], but correspond well with paramagnetic impurity models of the $Fe^{3+}$ ions sitting on the rare-earth site [18,19]. The number of resonances corresponds to the number of transitions possible between the six (2$S$+1) states of the $^6A_1$ spin multiplet ($L$=0, $S$=5/2) of the $Fe^{3+}$ ions in the crystal field. It is, however, unclear why these resonances are prevalent in the birefringence measurements and not in measurements of the Faraday effect.

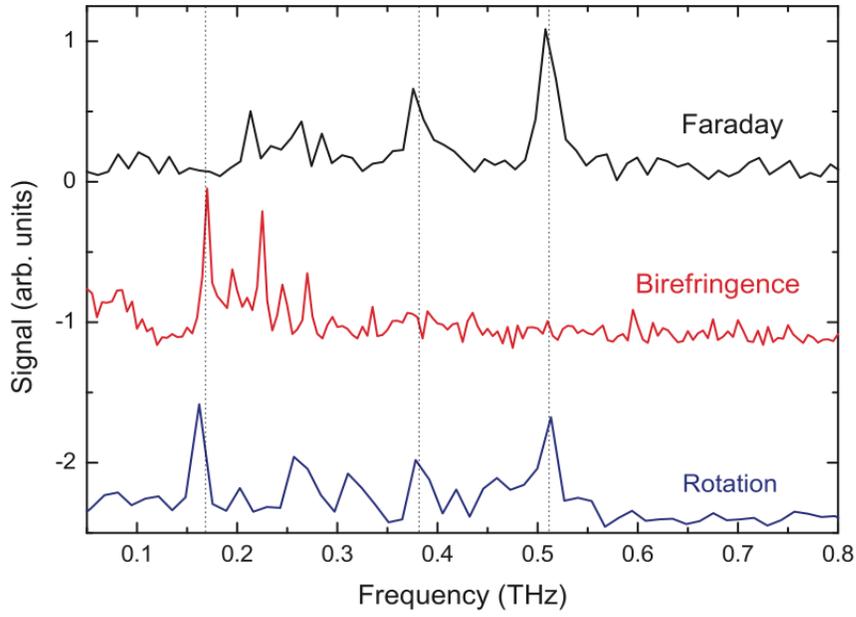

FIG. 3. (Color online) Different resonances appear in DyFeO$_3$ depending on whether the Faraday rotation or birefringence is measured. In Faraday measurements, the magnetic modes at 0.375 and 0.51 THz dominate, while time-resolved birefringence shows resonances in the range 0.15-0.30 THz. Measurements of optical rotation away from a minimum in the static birefringence generally contains both contributions.

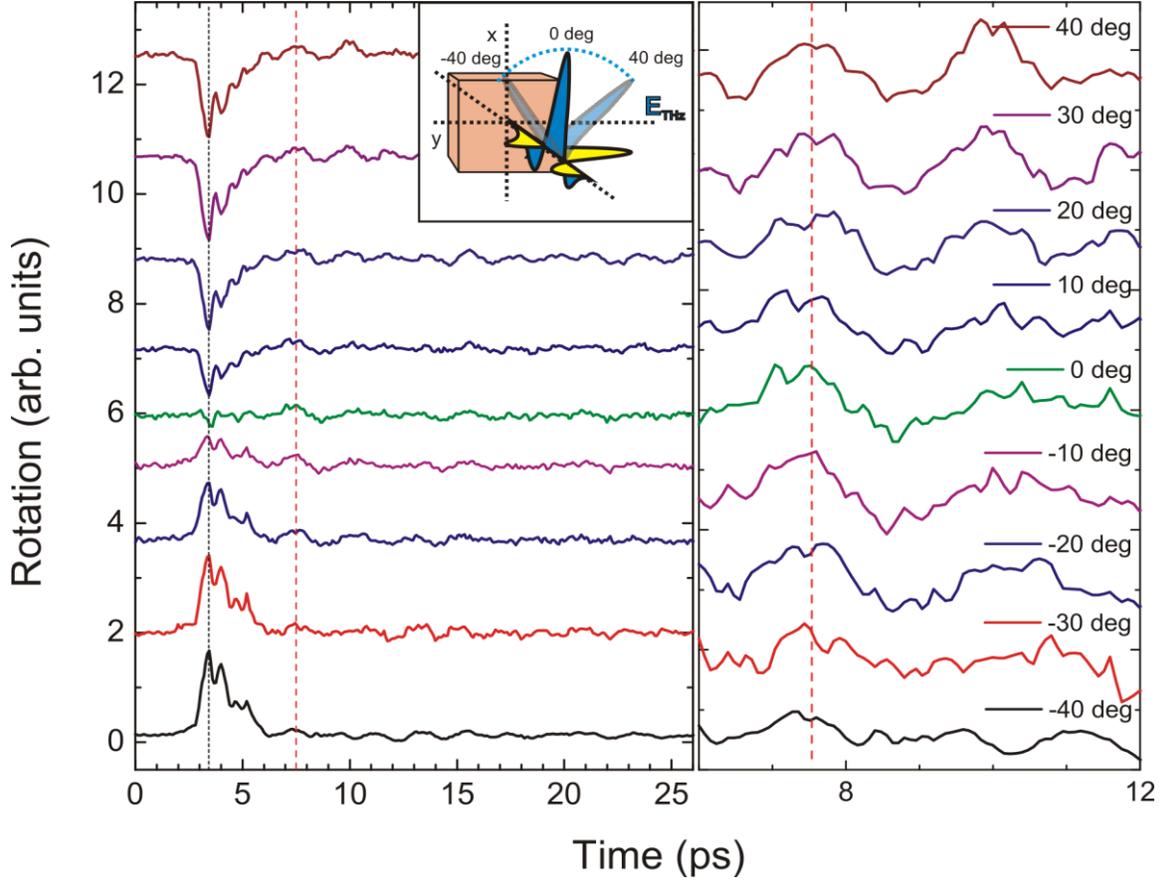

FIG. 4. (Color online) Measurements of the transient rotation as a function of the angle $\Theta$ of the $\mathbf{B_{THz}}$ pump field. The transient response at the pump-probe overlap shows strong dependence on the $\mathbf{B_{THz}}$-field angle, a fitting of this response shows a sinusoidal $\sin^2\theta$ behavior with an $180^0$ period. This is consistent with a modification of the optical susceptibility along the $\mathbf{B_{THz}}$-field direction. An examination of the subsequent coherent excitations, right, shows no systematic shift in the phase of these oscillations.

To further explore the nature of the THz pulse interaction with $DyFeO_3$, we studied the relation of the transient feature at time zero (pump-probe temporal overlap) and the subsequent oscillatory magnon signal. The polarization angle $\theta$ of the incident THz pulse was varied with respect to the in-plane *a* and *b* crystals axes, while the probe polarization and external field $B_{ext}$ were held in a fixed direction (Fig. 4). As the electric field of the THz pulse was rotated away from the polarization plane of the probe field, the transient birefringence signal appears at time zero. The signal's angular dependence is found to be sinusoidal with an $180^0$ period. Further measurements of the strength of this feature with THz power demonstrate that it is linearly proportional to the THz field. The symmetry of the $DyFeO_3$ crystal is orthorhombic and centrosymmetric with point group mmm (becoming m'm'm when the magnetic symmetry at 300 K is included). For this symmetry the pure first-order electro-optical effect is zero [20]. Two possibilities exist for the origin of this signal.

Either the THz pulse couples to the $Fe^{+3}$ impurity sites, or directly to the crystals magnetic birefringence when the permittivity tensor elements of the form $\varepsilon_{ii} = c_{iizx}M_zL_x$ can contribute, assuming that the THz field acts primarily on $M_z$.

In summary, we have demonstrated resonant excitation of both antiferromagnetic and ferromagnetic magnon modes at THz frequencies in dysprosium iron orthoferrite $DyFeO_3$. The dependence of the observed resonances on the applied external field $\mathbf{B_{ext}}$ unambiguously shows that the THz magnetic field $\mathbf{B_{THz}}$ component couples directly to the ferromagnetic moment. This has implications for future development of magnonic devices since it proofs the concept of manipulating magnons with ultrafast magnetic field pulses at very high frequencies.

**Acknowledgements**

This research was carried out at the Linac Coherent Light Source (LCLS) at SLAC National Accelerator Laboratory. LCLS is an Office of Science User Facility operated for the U.S. Department of Energy Office of Science by Stanford University. This work is supported by the Department of Energy, Office of Science, Basic Energy Sciences, Materials Sciences and Engineering Division, under Contract DE-AC02-76SF00515. R.V.P would like to acknowledge a partial support from project 14.B25.31.0025 by the Russian Ministry of Education and Science and RFBR project 15-02-04222a. We thank A.V. Kimel and R. Mikhaylovskiv for discussions.